\documentclass[sigconf]{acmart}

\usepackage{graphicx}
\usepackage{multirow}
\usepackage{subcaption}
\usepackage{booktabs} 

\usepackage{algorithm}
\usepackage{algorithmic}
\usepackage{hyperref}
\usepackage{listings}
\AtBeginDocument{%
  }

\copyrightyear{2025}
\acmYear{2025}
\setcopyright{acmlicensed}\acmConference[MM '25]{Proceedings of the 33rd ACM International Conference on Multimedia}{October 27--31, 2025}{Dublin, Ireland}
\acmBooktitle{Proceedings of the 33rd ACM International Conference on Multimedia (MM '25), October 27--31, 2025, Dublin, Ireland}
\acmDOI{10.1145/3746027.3755059}
\acmISBN{979-8-4007-2035-2/2025/10}

\acmSubmissionID{1876}

\begin{document}

\title{Neural Video Compression with In-Loop Contextual Filtering and Out-of-Loop Reconstruction Enhancement}

\author{Yaojun Wu}
\email{wuyaojun@bytedance.com}
\orcid{0000-0002-8138-4186}
\affiliation{%
  \institution{Bytedance China}
  \city{Beijing}
  \country{China}
}

\author{Chaoyi Lin}
\email{linchaoyi.cy@bytedance.com}
\orcid{0009-0002-7770-6821}
\affiliation{%
  \institution{Bytedance China}
  \city{Hangzhou}
  \state{Zhejiang}
  \country{China}
}

\author{Yiming Wang}
\email{isymwang@gmail.com}
\orcid{0000-0001-5683-909X}
\affiliation{%
  \institution{Hohai University}
  \city{Nanjing}
  \state{Jiangsu}
  \country{China}
}

\author{Semih Esenlik}
\email{semih.esenlik@bytedance.com}
\orcid{0009-0003-5573-0326}
\author{Zhaobin Zhang}
\email{zhaobin.zhang@bytedance.com}
\orcid{0000-0001-5961-5163}
\affiliation{%
  \institution{Bytedance Inc.}
  \city{San Diego}
  \state{CA}
  \country{USA}
}

\author{Kai Zhang}
\email{zhangkai.video@bytedance.com}
\orcid{0000-0002-4079-1797}
\author{Li Zhang}
\authornote{Corresponding Author.}
\email{lizhang.idm@bytedance.com}
\orcid{0000-0002-3463-9211}
\affiliation{%
  \institution{Bytedance Inc.}
  \city{San Diego}
  \state{CA}
  \country{USA}
}

\renewcommand{\shortauthors}{Yaojun Wu et al.}

\begin{abstract}
  This paper explores the application of enhancement filtering techniques in neural video compression. Specifically, we categorize these techniques into in-loop contextual filtering and out-of-loop reconstruction enhancement based on whether the enhanced representation affects the subsequent coding loop. In-loop contextual filtering refines the temporal context by mitigating error propagation during frame-by-frame encoding. However, its influence on both the current and subsequent frames poses challenges in adaptively applying filtering throughout the sequence. To address this, we introduce an adaptive coding decision strategy that dynamically determines filtering application during encoding. Additionally, out-of-loop reconstruction enhancement is employed to refine the quality of reconstructed frames, providing a simple yet effective improvement in coding efficiency. To the best of our knowledge, this work presents the first systematic study of enhancement filtering in the context of conditional-based neural video compression. Extensive experiments demonstrate a 7.71\% reduction in bit rate compared to state-of-the-art neural video codecs, validating the effectiveness of the proposed approach.
\end{abstract}




\maketitle

\begin{figure}[t]
\centering
\includegraphics[width=0.30\textwidth]{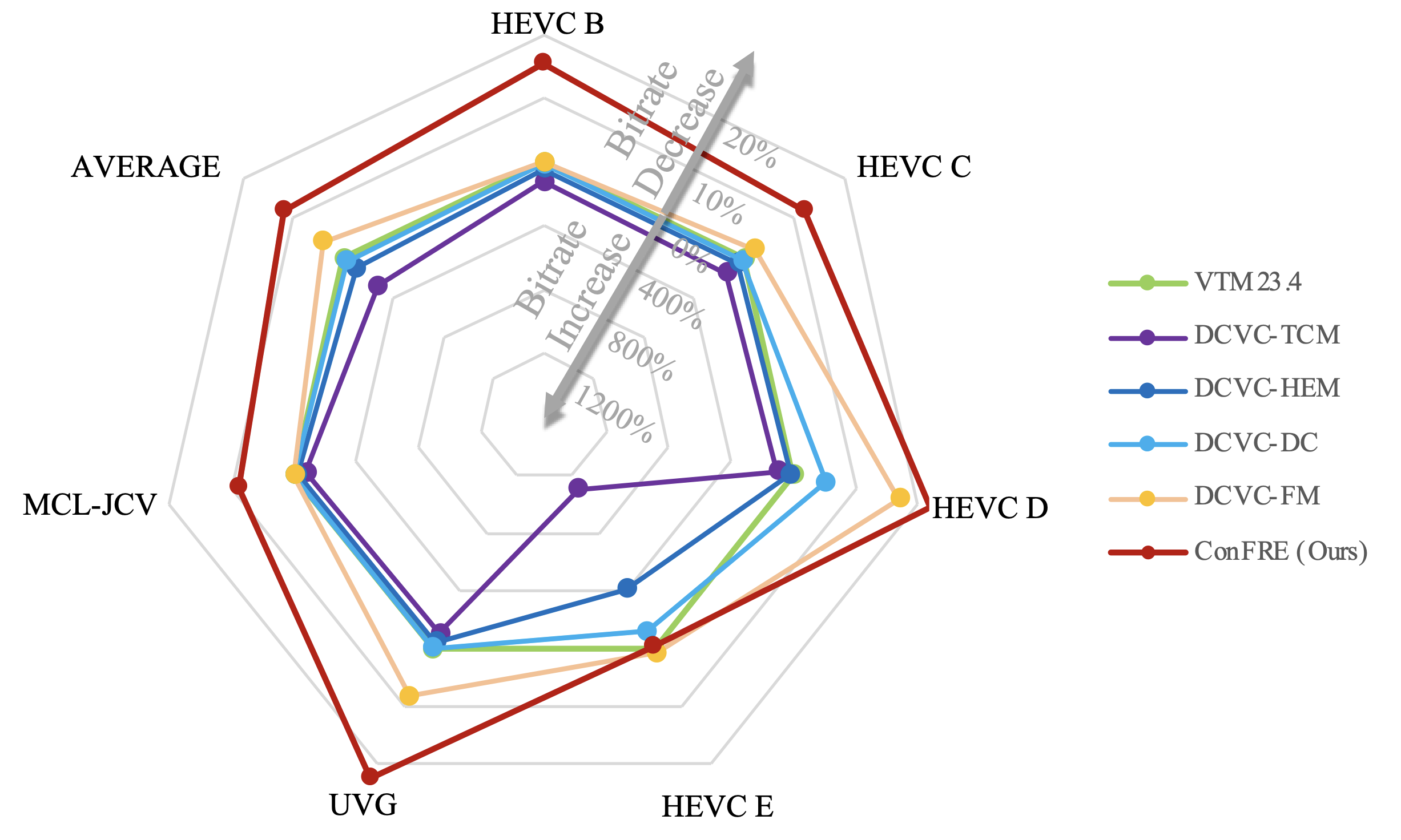}
\caption{BD-Rate comparison with H.266/VTM23.4~\cite{bross:2021overview} and state-of-the-art (SOTA) neural video compression methods, including DCVC-TCM~\protect\cite{sheng:2022temporal}, DCVC-HEM~\protect\cite{li:2022hybrid}, DCVC-DC~\protect\cite{li:2023neural}, and DCVC-FM~\protect\cite{li:2024neural}. The test condition follows a single-frame setting (intra period = -1), with all frames in the RGB color space.}
\label{fig:firstpage}
\end{figure}

\section{Introduction}
Traditional video coding~\cite{wiegand2003overview,sullivan2012overview,bross:2021overview} has long been essential for reducing transmission and storage costs. Over the past three decades, traditional coding techniques have achieved remarkable improvements in coding efficiency. However, further advancements have become increasingly challenging due to the rising algorithmic complexity and diminishing returns of handcrafted optimizations. Recently, neural-based compression has emerged as a transformative paradigm, leveraging deep learning to redefine video coding. This approach has demonstrated rapid progress, surpassing the performance of state-of-the-art traditional codecs~\cite{minnen2018joint,lieberman2023neural,yang2023lossy,li:2024neural} within a relatively short timeframe.

Early neural video compression (NVC) approaches adopted a residual coding framework~\cite{lu2019dvc}, closely resembling the structure of traditional video coding. Building on this framework, several efforts~\cite{agustsson2020scale,hu2022coarse} have been made to enhance its submodules. Later, conditional coding~\cite{li2021deep} was introduced and demonstrated superior performance in NVC. Unlike residual coding, which relies on predicted frames to reduce temporal redundancy, conditional-based NVC utilizes contextual feature information to store and propagate information from previously encoded frames. More recently, DCVC-FM~\cite{li:2024neural} introduced context refresh and long-sequence training, further alleviating error propagation accumulated during frame-by-frame coding. As a result, it achieves better performance than the state-of-the-art traditional codecs ECM~\cite{coban2025jvetaj} and VTM~\cite{bross:2021overview}.

Even though conditional-based NVC has made significant progress, several key areas for improvement remain unexplored. One open question is how to effectively integrate advanced filtering techniques into conditional NVC. Filtering, particularly in-loop filtering, is a promising approach for mitigating error propagation in long prediction chains and, in theory, offers substantial potential~\cite{karczewicz2021vvc}. However, leveraging these techniques effectively requires addressing critical challenges, such as balancing the trade-off between rate and distortion. Moreover, the potential benefits of out-of-loop enhancement in conditional NVC warrant further investigation.

Unlike NVC, filtering techniques are widely employed in traditional codecs. In-loop filters\cite{fu2012sample,karczewicz2021vvc} have been shown to be highly effective within traditional coding frameworks. However, applying filtering in conditional-based NVC introduces several unique challenges compared to traditional video coding frameworks. First, unlike traditional coding, where the coding framework is optimized in a modular fashion, NVC frameworks are optimized end-to-end, requiring careful consideration of both the placement and optimization objectives of the filtering process. Additionally, in contrast to in-loop filtering in traditional coding, the conditional-based NVC framework incorporates contextual information within the coding loop. While both contextual information and reconstruction information have the potential to enhance reconstruction, an open question remains: which type of information should be prioritized for enhancement within the coding loop?

To address the aforementioned issues, we introduce in-loop \textbf{Con}textual \textbf{F}iltering and out-of-loop \textbf{R}econstruction \textbf{E}nhancement (\textbf{ConFRE}) to improve the performance of conditional-based NVC. 
Rather than optimizing the reconstructed frame within the coding loop, we propose filtering contextual information as an alternative. This modification is motivated by two key factors. First, in NVC, most processing occurs in the feature domain, and contextual information is inherently more aligned with the feature domain than with the pixel domain. Second, utilizing the reconstructed frame at the beginning of the coding loop significantly extends the backpropagation path, potentially leading to unstable training or even model collapse.
For out-of-loop reconstruction enhancement, we propose enhancing the coded frame outside the coding loop. This design ensures that the enhanced frame does not interfere with subsequent coding operations, thereby enabling stable optimization and consistent improvements in reconstructed frame quality.
Building on the proposed filtering modules, we further introduce an encoder decision mechanism to determine whether filtering should be applied to the current frame. Instead of solely considering the rate-distortion performance of the current frame, the proposed decision mechanism evaluates the trade-off across all frames, which is crucial for the effective utilization of contextual filtering.
As shown in Figure~\ref{fig:firstpage}, our ConFRE framework achieves superior performance compared to previous state-of-the-art methods, demonstrating the effectiveness of the proposed approach.

The contributions of this work can be summarized as follows:
\begin{itemize} 
\item We propose an in-loop contextual filtering method to address the issue of error propagation. This approach is carefully engineered, with particular attention to key design factors that influence its performance.
\item Additionally, we introduce a simple yet highly effective out-of-loop filtering technique applied after the reconstruction of each frame. 
\item We propose an adaptive coding decision mechanism that intelligently controls each filter on and off, which is optimized to achieve the best rate-distortion (R-D) performance across the entire video sequence. 
\end{itemize}

\begin{figure*}[t]
\centering
\includegraphics[width=0.8\textwidth]{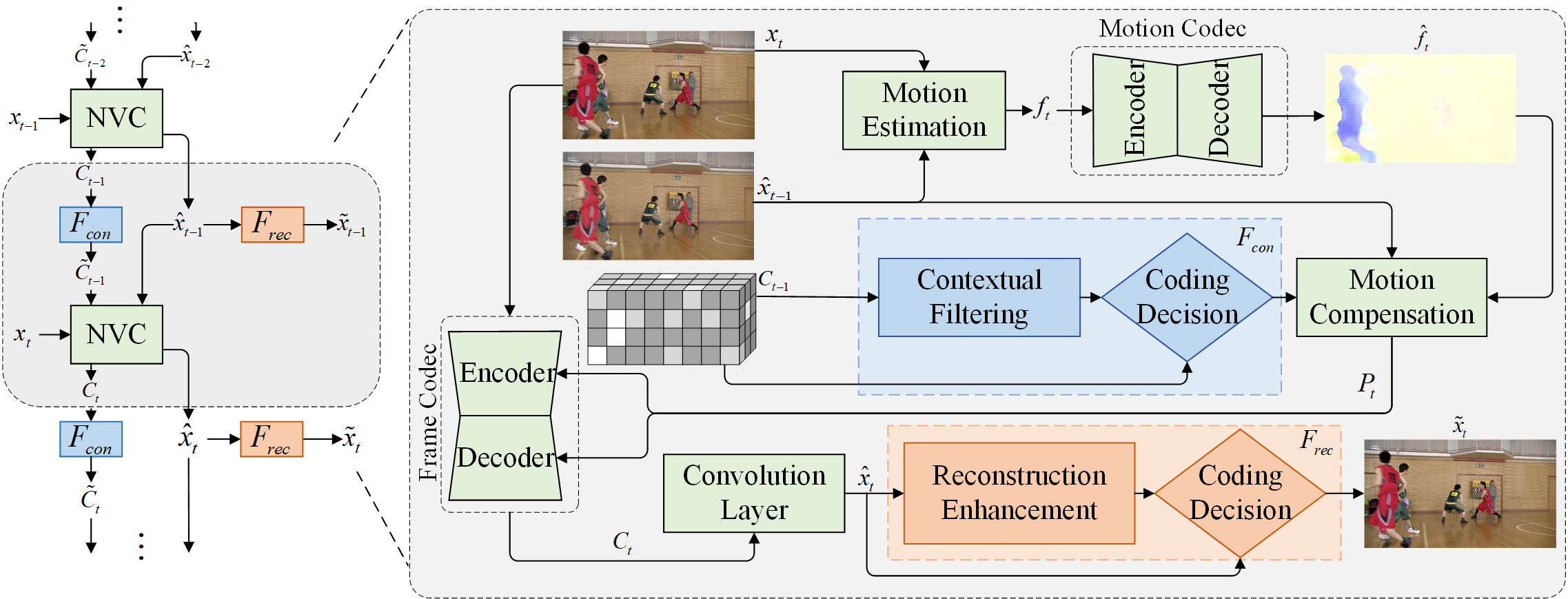}
\caption{Overall framework of the proposed ConFRE. $x_t$, $\hat{x}_t$, and $\tilde{x}_t$ denote the $t$-th frame, the reconstructed $t$-th frame, and the enhanced $t$-th frame, respectively. Similarly, $C_t$ and $\tilde{C}_t$ represent the $t$-th contextual frame and the enhanced $t$-th contextual frame. Modules enclosed in green boxes correspond to components from conditional NVC, while blue and orange boxes indicate the proposed in-loop contextual filtering ($F_{con}$) and out-of-loop reconstruction enhancement ($F_{rec}$), respectively.}
\Description[Overall framework of the proposed ConFRE.]{Overall framework of the proposed ConFRE. $x_t$, $\hat{x}_t$, and $\tilde{x}_t$ denote the $t$-th frame, the reconstructed $t$-th frame, and the enhanced $t$-th frame, respectively. Similarly, $C_t$ and $\tilde{C}_t$ represent the $t$-th contextual frame and the enhanced $t$-th contextual frame. Modules enclosed in green boxes correspond to components from conditional NVC, while blue and orange boxes indicate the proposed in-loop contextual filtering ($F_{con}$) and out-of-loop reconstruction enhancement ($F_{rec}$), respectively.}
\label{fig:framework}
\end{figure*}

\section{Related Work}

\subsection{Neural Video Compression}
Neural video compression was initially developed based on a residual coding framework, demonstrating superior performance over traditional codecs as early as 2019~\cite{chen2019learning,lu2019dvc}. Since then, extensive research has focused on enhancing submodules within this framework~\cite{djelouah2019neural,rippel2021elf,liu2023mmvc}. Notable advancements include improvements in motion estimation and compensation, such as scale-space optical flow~\cite{agustsson2020scale}, coarse-to-fine structures~\cite{hu2022coarse}, and multiscale motion compensation~\cite{liu2020neural}. Furthermore, multi-reference frame techniques~\cite{lin2020m} and block-based mode selection methods~\cite{liu2023mmvc} have also been explored.

Instead of directly subtracting the prediction frame from the current frame to remove temporal redundancy, conditional coding tends to maintain high-dimensional contextual feature information~\cite{liu2020conditional,ladune2021conditional,li2021deep,ho2022canf,qi2023motion}. This information is used as the conditional input in the transformation module to better utilize temporal correlations in the compression. Building upon this framework, DCVC-TCM~\cite{sheng:2022temporal} proposes a temporal context mining module to enhance the utilization of temporal context. Similarly, DCVC-HEM~\cite{li:2022hybrid} employs a spatio-temporal model in entropy probability modeling. Additionally, in DCVC-DC~\cite{li:2023neural}, offset diversity is introduced to enhance context diversity in both spatial and temporal dimensions, further boosting coding performance. Finally, DCVC-FM~\cite{li:2024neural} introduces feature modulation to simultaneously increase the dynamic rate range and mitigate error propagation. This approach suppresses traditional codecs and demonstrates the great potential of learned video compression. Despite the substantial progress in conditional coding, how to effectively integrate filtering techniques into neural video compression remains an open question, which motivates this study.

\subsection{Filtering in Traditional Video Compression}
Filtering techniques have been extensively studied in traditional video codecs~\cite{sullivan2012overview,chen2018overview,bross:2021overview}. For instance, Motion-Compensated Temporal Filtering (MCTF)~\cite{dubois1984noise} aligns blocks between reference frames and the current uncompressed frame to improve temporal consistency. Sample Adaptive Offset (SAO)~\cite{fu2012sample} mitigates sample distortion by classifying reconstructed samples and applying category-specific offsets to enhance their quality. Additionally, the deblocking filter~\cite{karczewicz2021vvc} is designed to reduce blocking artifacts, while Adaptive Loop Filters (ALF) and Cross-Component Adaptive Loop Filters further refine reconstructed frames through adaptive processing.

Inspired by the success of neural networks, several studies~\cite{liu2020deep,shao2023low} have explored enhancing filtering techniques in traditional video compression by leveraging deep learning. In~\cite{Li_2022_CVPR, shao2023low}, a learning-based loop filter is introduced to reduce compression artifacts in traditional codecs\cite{coban2025jvetaj}. Similarly,~\citet{Wang_2022_CVPR} propose an in-loop filter that integrates Generative Adversarial Networks (GANs) to improve compression performance, particularly from the perspective of subjective quality. Motivated by the promising potential of learned filtering, standardization groups have also begun exploring Neural Network-Based Video Coding (NNVC)~\cite{galpin2019jvetaj}. While filtering has significantly improved traditional coding, its integration into the conditional coding framework remains relatively unexplored.

\section{Proposed Method}

\subsection{Problem Formulation}
Our work builds upon the latest conditional-based NVC framework~\cite{li:2024neural}. Let $x_t$ denote the $t$-th frame to be encoded. The motion estimation network $g_{me}$~\cite{spynet2017}, parameterized by $\theta_{me}$, is first employed to estimate the optical flow $f_t$ between $x_t$ and the previously encoded frame $\hat{x}_{t-1}$. This optical flow $f_t$ is subsequently compressed through a series of operations, including a parametric analysis transformation $g_{ma}$ (parameterized by $\theta_{ma}$), quantization $Q$, and a parametric synthesis transformation $g_{ms}$ (parameterized by $\theta_{ms}$), resulting in the reconstructed optical flow $\hat{f}_t$. 
Next, the motion compensation network $g_{mc}$ generates the predicted conditional information $p_t$. Specifically, $g_{mc}$ takes the encoded frame $\hat{x}_{t-1}$, the compressed flow $\hat{f}_t$, and the contextual information $c_{t-1}$ from the previous frame as inputs. These inputs are processed within the motion compensation network through warping and feature extraction, guided by the parameters $\theta_{mc}$. 
The frame codec is then applied to compress the current frame’s information. This involves a parametric conditional analysis transformation $g_{fa}$ (parameterized by $\theta_{fa}$), followed by quantization $Q$, and a parametric conditional synthesis transformation $g_{fs}$ (parameterized by $\theta_{fs}$). The contextual information of the current frame, $c_t$, is further processed through a single convolutional layer $g_{conv}$, parameterized by $\theta_c$, to produce the reconstructed frame $\hat{x}_t$.
In summary, the entire coding process of conditional-based NVC can be summarized as follows:
\begin{align}
& f_t = g_{me}(x_t, \hat{x}_{t-1}; \theta_{me}), \\
& \hat{f}_t = g_{ms}(Q(g_{ma}(f_t;\theta_{ma})); \theta_{ms}), \\
& p_t = g_{mc}(\hat{x}_{t-1}, \hat{f}_t, c_{t-1}; \theta_{mc}), \label{eq:mc} \\
& c_t = g_{fs}(Q(g_{fa}(x_t, p_t; \theta_{fa})), p_t; \theta_{fs}), \\
& \hat{x}_t = g_{conv}(c_t; \theta_{c}).
\end{align}

Based on the structure of the conditional-based NVC, we propose contextual filtering $F_{con}$ and reconstruction enhancement $F_{rec}$ to further boost the coding performance, as illustrated in Figure~\ref{fig:framework}. The details of our solution will be discussed in the following sections.

\subsection{In-loop Contextual Filtering}

To address the issue of error propagation in long video sequences, ~\citet{li:2024neural} proposed the context refresh in the coding process, which periodically updates the contextual information to mitigate accumulated errors. Building on this strategy, we introduce an in-loop contextual filtering mechanism.

Rather than directly utilizing contextual information $c_{t-1}$ in motion compensation, we propose to refine it before usage. When contextual filtering is enabled, the motion compensation process in Equation.~(\ref{eq:mc}) can be reformulated as follows:
\begin{align}
& \tilde{c}_{t-1} = g_{cf}(c_{t-1}; \theta_{cf}), \\
& p_t = g_{mc}(\hat{x}_{t-1}, \hat{f}_t, \tilde{c}_{t-1}; \theta_{mc}),
\end{align}
where $g_{cf}$ represents the contextual filtering network, parameterized by $\theta_{cf}$. Unlike existing approaches that directly utilize $c_{t-1}$, our method incorporates the refined contextual information $\tilde{c}_{t-1}$ into the motion compensation process, thereby reducing error propagation and enhancing temporal consistency.

Align with conditional NVC, the training objective of contextual filtering is to enhance the quality of the current frame while minimizing bit rate consumption. This objective can be optimized using the Lagrange multiplier method, formulated as:
\begin{align}
L_{cf} & =  \frac{1}{T} \sum_{t=0}^T (R + \lambda_t \times D) \\
       & = \frac{1}{T} \sum_{t=0}^T (R(Q(g_{ma}(f_t;\theta_{ma}))) + \nonumber \\
       & R(Q(g_{fa}(x_t, p_t; \theta_{fa})))+ \lambda_t \times D(x_t, \hat{x}_{t})),
\end{align}
where $R$ is the rate of the features to be transmitted, and D is the distortion loss, measured as mean squared error (MSE) in our method. Following the approach in~\citet{li:2024neural}, we incorporate hierarchical quality optimization and long-sequence training to mitigate error propagation. To streamline the optimization process, we adopt a multi-stage training strategy, following the methodology of~\citet{sheng:2022temporal}.

\begin{figure}[t]
    \centering
    \begin{subfigure}[b]{0.4\textwidth}
        \centering
        \includegraphics[width=\textwidth]{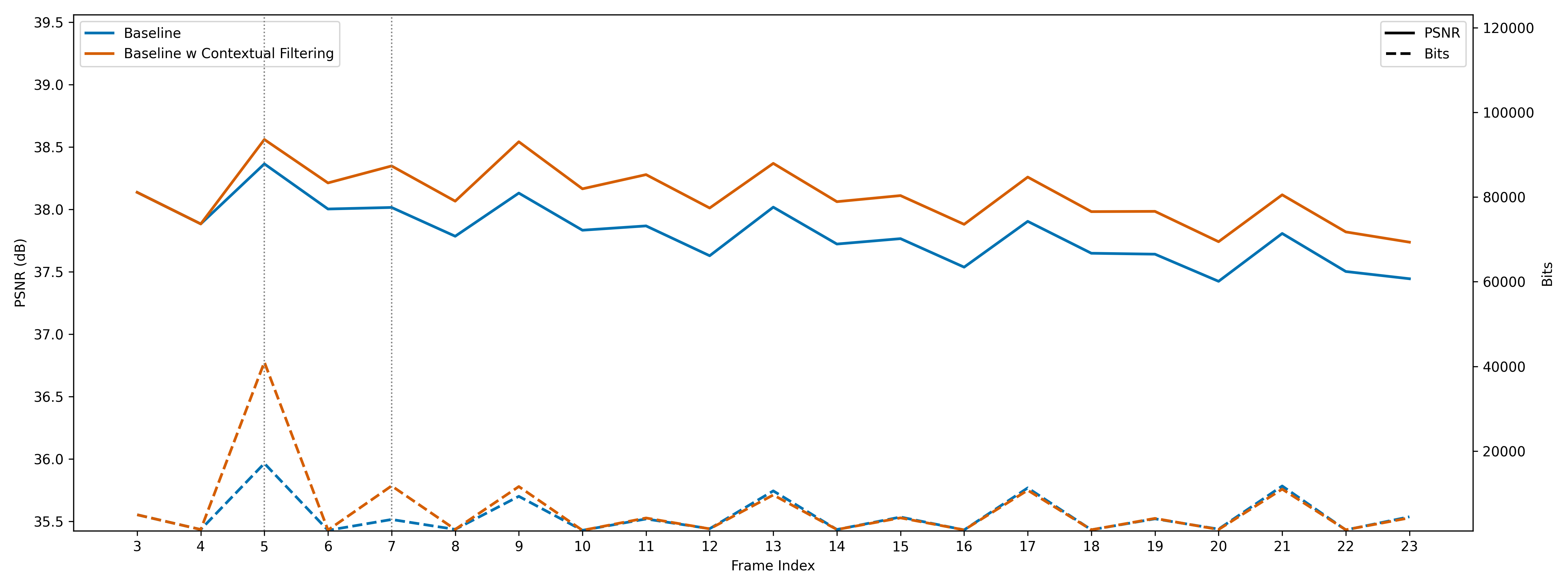}
        \caption{Result on KristenAndSara\_1280x720\_60 (Contextual filtering brings -4.35\% BD-rate improvement).}
        \label{fig:fourpeople}
    \end{subfigure}
    \hfill
    \begin{subfigure}[b]{0.4\textwidth}
        \centering
        \includegraphics[width=\textwidth]{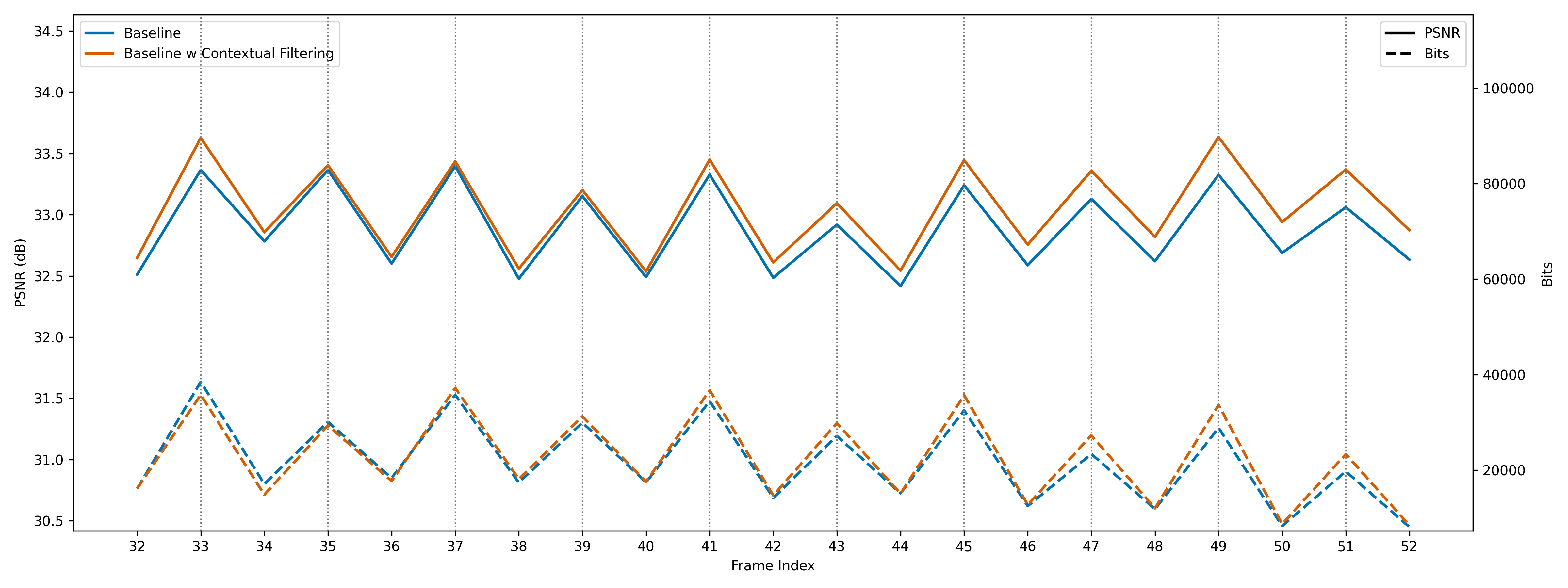}
        \caption{Result on BlowingBubbles\_416x240\_50 (Contextual filtering brings -4.04\% BD-rate improvement).}
        \label{fig:blowingbubbles}
    \end{subfigure}
    \caption{Example of rate-distortion changes after enabling contextual filtering. Gray vertical dash line means contextual filtering is enabled at this frame.}
    \Description[Example of rate-distortion changes after enabling contextual filtering]{Example of rate-distortion changes after enabling contextual filtering. Gray vertical dash line means contextual filtering is enabled at this frame.}
    \label{fig:inloop_vis}
\end{figure}

Figure~\ref{fig:inloop_vis} presents examples of performance variations observed when contextual filtering is enabled. The proposed contextual filtering method provides three key benefits:

\textbf{Improvement on Videos with Smooth Motion:} Figure~\ref{fig:fourpeople} presents an example from the HEVC E class, showcasing a video sequence with smooth motion characteristics. As illustrated in the figure, enhancing just a few frames in such video content can significantly elevate the quality of the entire sequence. While this enhancement may slightly increase the bit rate for certain frames, it ultimately leads to better overall rate-distortion performance.

\textbf{Replacement of Context Refresh:} As shown in Figure~\ref{fig:blowingbubbles} (Frame 33), previous approaches typically employed context refresh at this frame. However, when contextual filtering is applied, it serves as a substitute for context refresh, resulting in higher quality and a lower bit rate, not only for the current frame but also for subsequent frames.

\textbf{Quality Boost with Minimal Rate Increase:} As shown in Figure~\ref{fig:blowingbubbles} (Frames 40–52), contextual filtering improves the overall quality with only a marginal increase in bit rate. Given the rate-distortion trade-off in the current frame, enabling contextual filtering remains beneficial for enhancing coding efficiency.

\subsection{Out-of-loop Reconstruction Enhancement}
Instead of directly outputting the current coded frame $\hat{x}_t$, out-of-loop reconstruction enhancement further improves coding performance beyond the coding loop by employing a reconstruction enhancement network $g_{re}$, formulated as:
\begin{align}
\tilde{x}_t = g_{re}(\hat{x}_t; \theta_{re}).
\end{align}

Since the enhanced frame is only utilized to improve the quality of the current frame and does not influence the subsequent coding process, the optimization objective of reconstruction enhancement is straightforward, given by:
\begin{align}
L_{re} = D(x, \tilde{x}_t),
\end{align}
where $D$ represents the distortion loss, computed as mean squared error (MSE) in our method. We employ a training augmentation strategy that incorporates random frame selection and variable-rate training. During each training iteration, we randomly sample a rate point and a frame at different temporal positions, allowing the network to learn how distortions evolve over time and across different compression rates, thereby optimizing enhancement quality under diverse coding conditions.

\begin{figure}[t]
\centering
\includegraphics[width=0.45\textwidth]{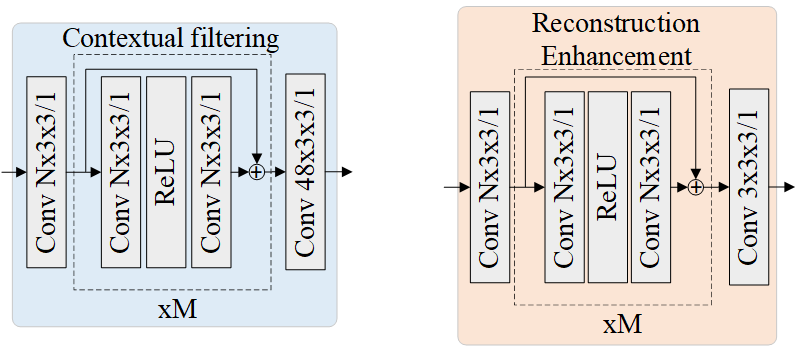}
\caption{Structure of the contextual filtering and reconstruction enhancement. The left panel illustrates the architecture of the contextual filtering module, while the right panel presents the reconstruction enhancement module. Convolution parameters are denoted as: number of filters $\times$ height of kernel $\times$ width of kernel $/$ stride.}
\label{fig:structurefilter}
\end{figure}

\subsection{Detailed Structure of the Proposed Modules}
\label{sec:structure}
Figure~\ref{fig:structurefilter} illustrates the architecture of the contextual filtering and reconstruction enhancement modules, which share a unified design. The input is first projected into the feature space via a $3\times3$ convolutional layer with $N$ filters. This is followed by $M$ residual blocks, each consisting of two $3\times3$ convolutional layers with intermediate ReLU activations, to refine the feature representations. Finally, a concluding $3\times3$ convolutional layer, symmetric to the initial embedding layer, maps the features back to the target space, producing either the enhanced contextual representation or the improved reconstructed frame.

The design of this module is guided by three key principles: simplicity, to ensure computational efficiency; scalability, to support diverse application scenarios; and stability, during both training and inference. To achieve these goals, we adopt the widely used residual structure as the fundamental building block, avoiding more complex architectures such as transformers or attention modules, which typically incur higher computational overhead.

To enable smooth transitions between the pixel/feature domains and the target representation space, we introduce shallow convolutional layers before and after the residual stack. This simple yet effective design facilitates better representation learning while maintaining efficiency.

Moreover, the model’s complexity can be flexibly adjusted by tuning the hyperparameters $M$ and $N$, making it well-suited for deployment under varying system constraints. In our main configuration, we set $M=8$ and $N=32$.

\subsection{Adaptive Coding Decision During Encoding}

To ensure that contextual filtering and reconstruction enhancement contribute positively to the overall compression performance, we propose an adaptive coding decision mechanism. The detailed procedure is presented in Algorithm~\ref{alg:algorithm}.

While contextual filtering improves reconstruction quality, it may also increase the bitstream size, potentially leading to suboptimal rate-distortion (R-D) trade-offs. To address this, our strategy dynamically determines whether to enable filtering on a per-frame basis, aiming to balance distortion and rate across the entire sequence. This decision process is guided by two core observations:

\textbf{Intra-Period Reference Dependencies.} Following the context refresh scheme in~\cite{li:2024neural}, the encoding sequence is divided into fixed-length periods (e.g., 32 frames), with each period starting with a context refresh. Within a period, early frames act as reference points for subsequent frames. Enhancing these early frames—even at the cost of higher bitrates—often leads to improved overall performance due to the propagation of higher-quality references.

\textbf{Inter-Period Global Dependencies.} Beyond a single refresh period, earlier frames in the entire sequence influence a larger number of subsequent frames. As encoding proceeds sequentially, the quality of earlier frames has a compounding effect on later predictions. Therefore, investing bitrate in these frames can yield long-term benefits, while frames near the end of the sequence are less impactful and may not warrant additional filtering.

\begin{algorithm}[t]
    \caption{Encoding with Adaptive Coding Decision}
    \label{alg:algorithm}
    \textbf{Input}: Current Frame $x_t$, Contextual information $c_{t-1}$, Previous coded frame $\hat{x}_{t-1}$, Frame number $t$ \\
    \textbf{Parameters}: Context refresh period $crp$, Maximum quality counter $mqc$, Progressive factor $pf$, Total frame length $tfl$, Contextual counter $con$ \\
    \textbf{Output}: Bitstream $b_t$, flag for contextual filtering $f_{cf}$, flag for reconstruction enhancement $f_{re}$, reconstruction $\tilde{x}_{t}$
    \begin{algorithmic}[1] 
        \STATE Perform contextual filtering: $\tilde{c}_{t-1} = g_{cf}(c_{t-1}; \theta_{cf})$
        \IF{$t\%crp==0$}
        \STATE $c_{t-1} = c_{t-1} * 0$
        \ENDIF
        \STATE Encode to get bitstream $b_{t1}$ and Reconstruction $\hat{x}_{t1}$ with $c_{t-1}$: $b_{t1}, \hat{x}_{t1}=Encode(x, c_{t-1}, \hat{x}_{t-1})$
        \STATE Encode to get bitstream $b_{t2}$ and Reconstruction $\hat{x}_{t2}$ with $\tilde{c}_{t-1}$: $b_{t2}, \hat{x}_{t2}=Encode(x, \tilde{c}_{t-1}, \hat{x}_{t-1})$
        \STATE $r_1, d_1 = len(b_{t1}), mse(\hat{x}_{t1}, x_t)$
        \STATE $r_2, d_2 = len(b_{t2}), mse(\hat{x}_{t2}, x_t)$
        \IF {$d_2<d_1$ and ($con<mqc$ or $L_{pl} < 0$) }
        \STATE $con = con+1$
        \STATE $b_t = b_{t2}$, $\hat{x}_t = \hat{x}_{t_2}$, $d = d_2$, $f_{cf}=1$
        \ELSE
        \STATE $b_t = b_{t_1}$, $\hat{x}_t = \hat{x}_{t_1}$, $d = d_1$, $f_{cf}=0$
        \ENDIF
        \IF{$t\%crp==0$}
        \STATE $con=0$
        \ENDIF
        \STATE Perform reconstruction enhancement: $\tilde{x}_t=g_{re}(\hat{x}_t; \theta_{re})$
        \IF{$d<mse(\tilde{x}_t, x)$}
        \STATE $\tilde{x}_t = \hat{x}_t$, $f_{re}=0$
        \ELSE
        \STATE $f_{re}=1$
        \ENDIF
        \STATE \textbf{return} $b_t$, $f_{cf}$, $f_{re}$, $\tilde{x}_{t}$
    \end{algorithmic}    
\end{algorithm}

\begin{figure*}[h]
    \centering
    \includegraphics[width=1.0\textwidth]{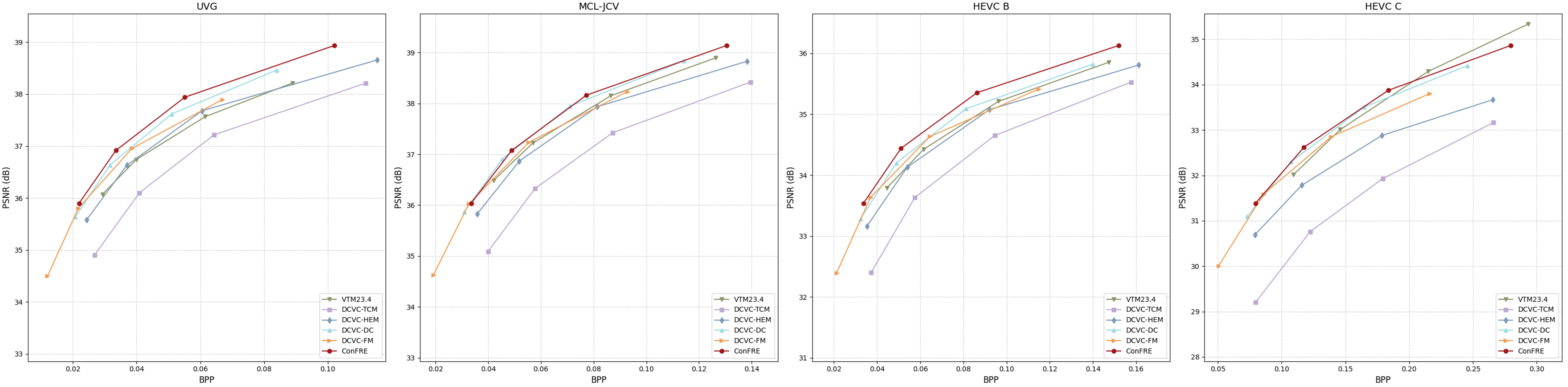}
    \vspace{-2em}
    \caption{Rate-distortion curves on UVG, MCL-JCV and HEVC datasets. Test condition is 96 frames with intra period=32. The quality indexes of DCVC-FM are set to match the bit-rate range of DCVC-DC.}
    \vspace{-1em}
    \label{fig:F96G32}
\end{figure*} 

\begin{table*}[t]
    \centering
    \caption{BD-Rate comparison in RGB colorspace. Test condition is 96 frames with intra period=32.}
    \vspace{-1em}
    \begin{tabular}{cccccccc}
        \toprule
                  &  HEVC B & HEVC C & HEVC D & HEVC E & UVG    & MCL-JCV & Average \\
        \midrule
        VTM-23.4  & 0.00\%  & 0.00\% & 0.00\%  & 0.00\% & 0.00\%  & 0.00\%  & 0.00\%  \\
        DCVC-TCM  & 35.81\% & 71.65\%& 32.21\% & 91.01\%& 28.84\% & 41.39\% & 50.15\% \\
        DCVC-HEM  & 4.29\%  & 28.13\%& -3.30\% & 28.72\%&-4.69\%  & 5.94\%  & 9.85\%  \\
        DCVC-DC   & -9.01\% & -3.54\%&-24.99\% & -10.29\%&-17.96\%&-9.03\%  &-12.47\% \\
        DCVC-FM   & -1.69\% & 4.63\% &-19.40\% & -8.58\% &-11.40\% &-0.53\% &-6.16\%  \\
        \midrule
        \textbf{ConFRE}    & \textbf{-17.59\%}& \textbf{-4.60\%} &  \textbf{-21.16\%} & \textbf{-9.13\%}  & \textbf{-24.18\%} & \textbf{-9.79\%} & \textbf{-14.41\%} \\
        \bottomrule
    \end{tabular}
    \label{tab:exp9632}
\end{table*}

Based on these insights, we design an adaptive decision strategy that jointly considers both local and global reference relationships. Specifically, we introduce a contextual counter $con$ that tracks the number of frames using contextual filtering within a refresh period. A maximum quality counter $mqc$ is used in conjunction with $con$: if $con < mqc$, filtering is enabled for frames that yield any quality gain. This guarantees that at least $mqc$ frames benefit from filtering within each period, provided such filtering is beneficial.

For the remaining frames, we apply a progressive rate-loss strategy to determine whether filtering should be applied. The decision criterion is defined as:

\begin{align}
L_{pl} = \frac{r_2 - r_1}{r_1} - pf \left( 1 - \frac{t}{tfl} \right),
\label{eq.lpl}
\end{align}

where $r_1$ and $r_2$ are the bitrates without and with contextual filtering, respectively, $t$ is the index of the current frame, $tfl$ is the total frame length, and $pf$ is a progressive factor that controls the maximum allowable rate increase. According to Equation.~\ref{eq.lpl}, contextual filtering is enabled only if $L_{pl} < 0$. In our implementation, $mqc$ and $pf$ are set to 2 and 0.16, respectively.

In addition, we also apply adaptive decision-making to the reconstruction enhancement module. Since this module does not affect the bitstream or subsequent frames, its application is determined solely by whether it improves the current frame’s reconstruction quality.

While our method relies on empirical thresholds and heuristic comparisons, it offers a practical and effective approximation to globally optimized R-D performance. It is simple to implement, content-aware, and easily integrates into real-time encoding pipelines with minimal overhead.

\section{Experimental Result}


\subsection{Experimental Settings}

\textbf{Datasets.} Following~\citet{li:2024neural}, we download raw Vimeo videos~\cite{anchen2021toflow} and preprocess them using scene detection and data cleaning techniques. This process results in a final training dataset consisting of 67,334 video sequences. For evaluation, we test the model on three benchmark datasets: HEVC B–E~\cite{bossen2013common}, UVG~\cite{mercat:2020uvg}, and MCL-JCV~\cite{wang:2016mcl}. All datasets are evaluated at their original resolutions.

\textbf{Training Conditions.} Our model is built upon the DCVC series~\cite{li:2023neural,li:2024neural}. We replicated the training process of these models and optimized it for RGB input, resulting in our baseline model, 
which serves as the foundation for subsequent experiments. Based on this baseline, we trained models for contextual filtering and reconstruction enhancement. All models were trained on four Tesla A100-80G GPUs. During each iteration, video sequences were randomly cropped into $256 \times 256$ patches without explicit downsampling. The batch size was set to 4 for contextual filtering and 16 for reconstruction enhancement.

\textbf{Test Conditions.} All evaluations were conducted under the low-delay setting, meaning only past frames (no B-frames) are used for the compression of the current frame. To assess compression efficiency, we use the Bjøntegaard Delta Rate (BD-rate)\cite{gisle2001calculation}, where negative values indicate bit rate savings, and positive values indicate an increase in bit rate. For comparison, we evaluate performance against the traditional codec H.266/VTM23.4\cite{bross:2021overview}. Additionally, we compare our results with existing NVC-based methods, including DCVC-TCM~\cite{sheng:2022temporal}, DCVC-HEM~\cite{li:2022hybrid}, DCVC-DC~\cite{li:2023neural}, and DCVC-FM~\cite{li:2024neural}. We compare these methods under different intra-periods (32 and -1) and frame lengths (96 and all frames) to verify the effectiveness of our proposed method under various settings. To align with the settings of most DCVC models, all evaluations are conducted in the RGB domain.

\begin{table}[t]
\centering
\caption{Ablation Results with Component Activation. The baseline model is utilized as the anchor in BD-rate calculation. "CF" denotes contextual filtering, and "RE" stands for reconstruction enhancement.}
\begin{tabular}{cccc}
\toprule
\textbf{Baseline} & \textbf{CF} & \textbf{RE} & \textbf{BD rate} \\
\midrule
\checkmark & & & 0.00\% \\ \hline
\checkmark & \checkmark & & -2.43\% \\ \hline
\checkmark & & \checkmark & -4.55\% \\ \hline
\checkmark & \checkmark & \checkmark & -6.04\% \\ 
\bottomrule
\end{tabular}
\label{tab:ablation}
\end{table}

\begin{figure*}[t]
    \centering
    \includegraphics[width=1.0\textwidth]{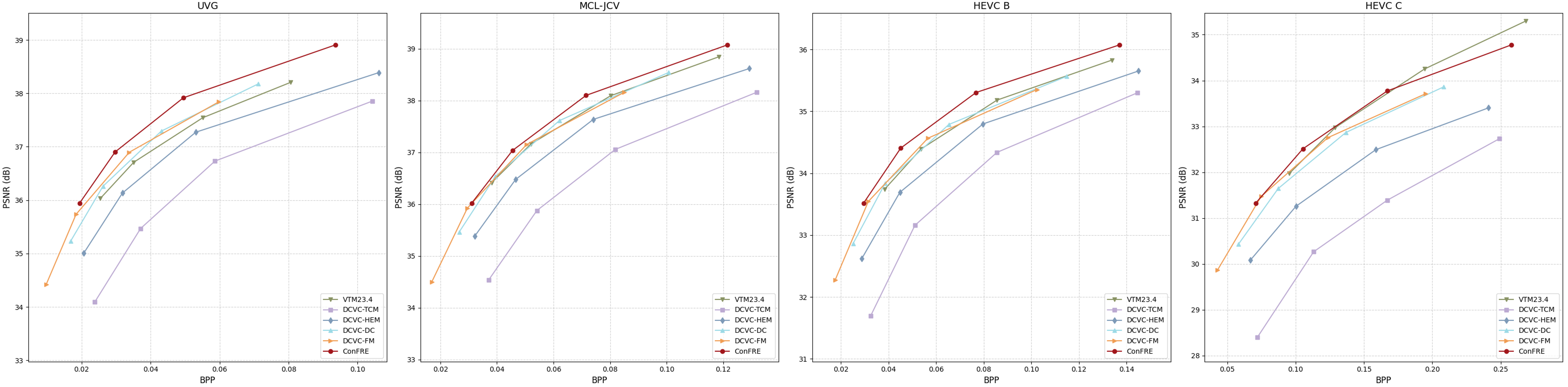}
    \vspace{-2em}
    \caption{Rate-distortion curves on UVG, MCL-JCV, and HEVC B and HEVC C datasets. Test condition is 96 frames with intra period=-1.}
    \Description[Rate-distortion curves on UVG, MCL-JCV, and HEVC datasets.]{}
    \label{fig:all_images96}
\end{figure*} 

\begin{table*}
    \centering
    \caption{BD-Rate comparison in RGB colorspace. Test condition is 96 frames with intra period=-1.}
    \vspace{-1em}
    \begin{tabular}{cccccccc}
        \toprule
                  &  HEVC B & HEVC C  & HEVC D    & HEVC E   & UVG      & MCL-JCV & Average  \\
        \midrule
        VTM-23.4  & 0.00\%  & 0.00\%  &  0.00\%   & 0.00\%   & 0.00\%   & 0.00\%  &  0.00\%  \\
        DCVC-TCM  & 62.87\% & 109.66\% &  52.34\%  & 270.25\% & 64.32\%  & 63.60\% &  103.84\% \\
        DCVC-HEM  & 19.18\% & 47.01\% &  6.30\%  & 107.04\% & 14.68\%  & 17.32\% &  35.26\% \\
        DCVC-DC   & -2.15\% & 10.61\% &  -19.12\%  & 12.96\% & -9.15\%   & -1.87\%  &  -1.45\% \\
        DCVC-FM   & 0.40\%  & 8.02\% &  -20.31\% & \textbf{-4.46\%}  & -11.04\%  & 0.03\%  & -4.56\%  \\
        \midrule
        \textbf{ConFRE}    & \textbf{-16.61\%}& \textbf{-0.88\%} &  \textbf{-21.70\%} & -0.33\%  & \textbf{-24.46\%} & \textbf{-9.31\%} & \textbf{-12.22\%}        \\
        \bottomrule
    \end{tabular}
    \label{tab:exp9696}
\end{table*}

\subsection{Comparison with Previous SOTA Methods}
\begin{figure*}[t]
    \centering
    \includegraphics[width=0.95\textwidth]{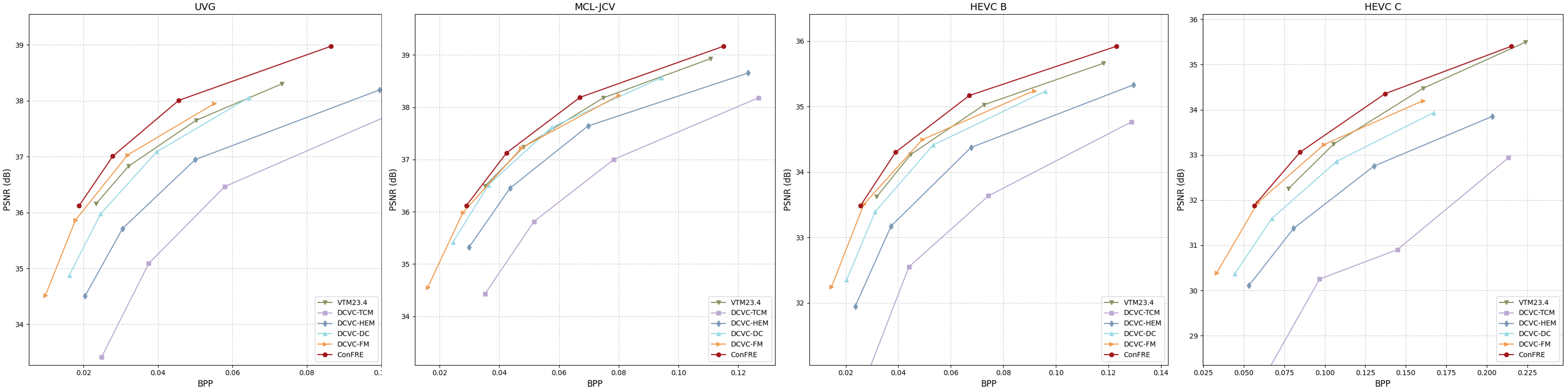}
    \vspace{-1em}
    \caption{Rate-distortion curves on UVG, MCL-JCV, and HEVC B and HEVC C datasets. Test condition is all frames with intra period=-1.}
    \label{fig:all_images}
    \Description[Rate-distortion curves on UVG, MCL-JCV, and HEVC datasets.]{Rate-distortion curves on UVG, MCL-JCV, and HEVC datasets. Test condition is all frames with intra period=-1.}
\end{figure*}

\textbf{Objective Quality.} 
Following the evaluation protocol of \citet{li:2024neural}, we firstly assess our model using 96-frame with an intra period of 32. In this test, all proposed modules are enabled in our method. The results, presented in Figure~\ref{fig:F96G32} and Table~\ref{tab:exp9632}, show that our method achieves a 14.41\% bitrate reduction compared to VTM 23.4 and an 8.25\% coding gain over DCVC-FM.
We further evaluate our model under a 96-frame, where the intra period is set to -1. As shown in Figure~\ref{fig:all_images96} and Table~\ref{tab:exp9696}, our approach consistently outperforms baselines, achieving a 12.22\% bitrate reduction over VTM 23.4 and a 7.66\% reduction over DCVC-FM.
In addition, Figure~\ref{fig:all_images} and Table~\ref{tab:exp-1-1} report results for a all-frames configuration with an intra period of -1. The proposed ConFRE method delivers the best compression performance among all compared approaches, yielding an 11.87\% bitrate reduction relative to VTM 23.4 and a 7.71\% reduction compared to DCVC-FM. Notably, this setting corresponds to the low-delay configuration widely adopted in practical video coding scenarios, further validating the effectiveness and applicability of our method.

\begin{figure*}[t]
\centering
\includegraphics[width=0.85\textwidth]{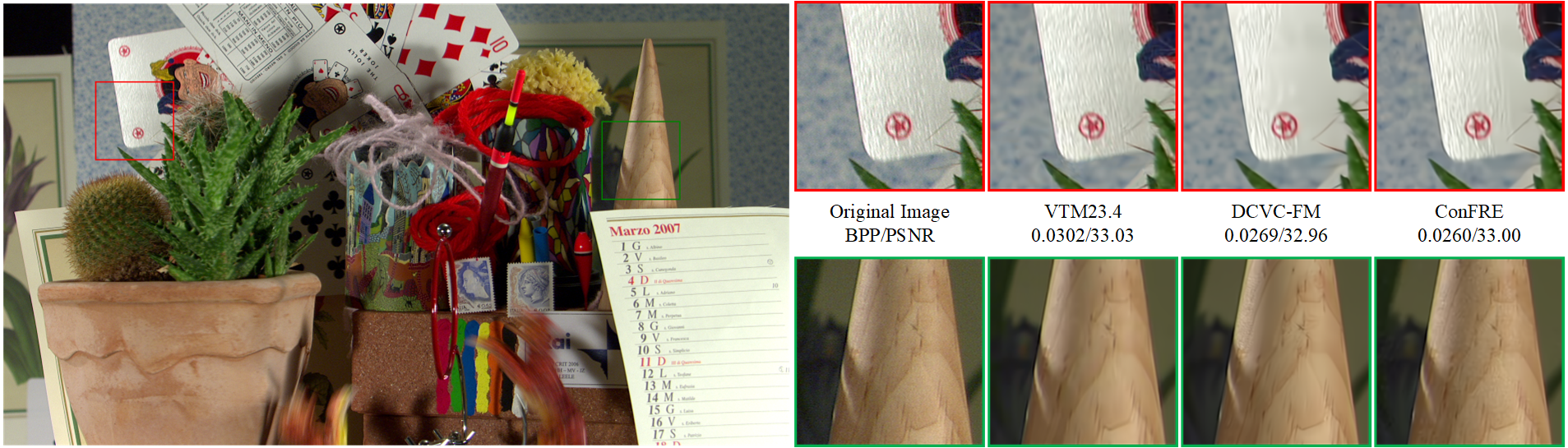}
\vspace{-1em}
\caption{Visual comparison. Test condition is all frames with intra period=-1. Compared to DCVC-FM, our solution demonstrates better texture retention, particularly in details like poker cards and wooden surfaces.}
\Description[Visual comparison. Test condition is all frames with intra period=-1.]{}
\label{fig:subjective}
\end{figure*}

\textbf{Subjective Quality.} Figure~\ref{fig:subjective} presents visual comparisons. Compared to DCVC-FM, our method exhibits superior texture retention across a wide range of visual details, leading to lower bit cost and higher PSNR. 

\begin{table}[t]
    \centering
    \caption{Sensitivity study of $mqc$ and $pf$. $mqc=2$ and $pf=0.16$ is utilized as the anchor. Result is tested on HEVC datasets. Test condition is all frames with intra period=-1.}
    \vspace{-1em}
    \begin{tabular}{cccccc}
        \toprule
        \textbf{$mqc$}      & 0  & 1 & 2 & 3 & 4 \\
        \midrule
        \textbf{BD-rate}    & 0.43\%  & 0.92\% & 0.00\% & 0.32\% & 1.12\% \\    
        \bottomrule
        \textbf{$pf$}       & 0.00  & 0.08 & 0.16 & 0.24 & 0.32 \\
        \midrule
        \textbf{BD-rate}    & 0.25\%  & 0.24\% & 0.00\% & 0.49\% & 0.93\% \\ 
        \bottomrule
    \end{tabular}
    \vspace{-1.2em}
    \label{tab:abconpf}
\end{table}

\subsection{Ablation Study}

\textbf{Ablation Study on the Proposed Modules.}
Table~\ref{tab:ablation} summarizes the individual contributions of each proposed module to the overall performance.
We first examine the effect of contextual filtering. By improving the quality of the current frame, this module effectively mitigates error propagation to subsequent frames, resulting in a 2.43\% coding gain.
Next, we evaluate the impact of reconstruction enhancement, which improves the reconstruction quality without increasing the bitrate, leading to a 4.55\% coding gain. This confirms its effectiveness in optimizing compression efficiency.
When both modules are integrated, the overall coding gain reaches 6.04\%, indicating that contextual filtering and reconstruction enhancement offer complementary benefits and jointly contribute to substantial improvements in compression performance.

\textbf{Ablation Study on $mqc$ and $pf$.}
In our adaptive coding decision mechanism for contextual filtering, two empirical parameters—$mqc$ and $pf$—are used to determine whether contextual filtering should be enabled. In this section, we analyze the influence of these parameters on overall performance. As shown in Table~\ref{tab:abconpf}, the best results are achieved when $mqc = 2$ and $pf = 0.16$, demonstrating the effectiveness of these settings in guiding the adaptive filtering process.

\begin{table}[t]
\centering
\caption{The time profile of the encoding and decoding procedures is shown, with results tested on a 1080p sequence. "NVC" refers to the NVC module, "AC" represents arithmetic coding, "CF" denotes contextual filtering, and "RE" stands for reconstruction enhancement.}
\vspace{-1em}
\resizebox{\linewidth}{!}{ 
\begin{tabular}{@{}ccccccc@{}}
\toprule
\multirow{2}{*}{\textbf{Item}} & \multirow{2}{*}{\textbf{Params(M)}} & \multirow{2}{*}{\textbf{Flops(G)}}&\multicolumn{2}{c}{\textbf{Encoder}} & \multicolumn{2}{c}{\textbf{Decoder}} \\
\cmidrule(lr){4-5} \cmidrule(lr){6-7} & & & \textbf{Time(ms)} & \textbf{Ratio} & \textbf{Time(ms)} & \textbf{Ratio} \\ 
\midrule
NVC                          & 19.78 & 2786.48 & 586.89  & 82.42\% & 254.09 & 87.94\% \\
AC                           & - & - & 62.39   & 8.76\%  & 1.97   & 0.68\%  \\
CF                           & 0.18 & 382.21 & 32.15  & 4.52\% & 2.28   & 0.79\%  \\
RE                           & 0.16 & 328.46 & 30.61  & 4.30\% & 30.61 & 10.59\% \\
\midrule
Total                        & 20.12 & 3497.15 & 712.04  & 100.00\% & 288.95 & 100.00\% \\
\bottomrule
\end{tabular}
}
\label{tab:time_costs}
\end{table}

\subsection{Complexity Analysis}
To evaluate the computational overhead of the proposed method, we conducted a detailed runtime profiling, as summarized in Table~\ref{tab:time_costs}.

\begin{table*}
    \centering
    \caption{BD-Rate comparison in RGB colorspace. Test condition is all frames with intra period=-1.}
    \vspace{-1em}
    \begin{tabular}{cccccccc}
        \toprule
                  &  HEVC B & HEVC C  & HEVC D    & HEVC E   & UVG      & MCL-JCV & Average  \\
        \midrule
        VTM-23.4  & 0.00\%  & 0.00\%  &  0.00\%   & 0.00\%   & 0.00\%   & 0.00\%  &  0.00\%  \\
        DCVC-TCM  & 125.41\% & 143.62\% &  99.22\%  & 1106.25\% & 107.02\%  & 77.26\% &  276.46\% \\
        DCVC-HEM  & 42.26\% & 47.66\% &  19.55\%  & 410.13\% & 46.42\%  & 23.30\% &  98.24\% \\
        DCVC-DC   & 10.21\% & 17.17\% &  -5.22\%  & 119.52\% & 7.62\%   & 1.68\%  &  25.16\% \\
        DCVC-FM   & 1.00\%  & -2.14\% &  -17.03\% & \textbf{-0.68\%}  & -8.30\%  & 2.22\%  & -4.16\%  \\
        \midrule
        \textbf{ConFRE}    & \textbf{-15.50\%}& \textbf{-11.92\%} &  \textbf{-22.05\%} & 9.62\%  & \textbf{-22.41\%} & \textbf{-8.98\%} & \textbf{-11.87\%}        \\
        \bottomrule
    \end{tabular}
    \label{tab:exp-1-1}
\end{table*}

For contextual filtering, the encoding process incurs a 4.52\% increase in encoding time and a 0.79\% increase in decoding time, both of which are relatively minor—especially on the decoder side. 
From a model complexity perspective, contextual filtering introduces only 0.18 million parameters and 382.21 GFLOPs, which is lightweight relative to the backbone network (NVC), which has 19.78 million parameters and 2786.48 GFLOPs. 

For reconstruction enhancement, the encoding time increases by 4.3\%, while the decoding time rises by 10.59\%. 
As the parameters and Flops, this module contains only 0.16 million parameters and contributes 328.46 GFLOPs. Nevertheless, its impact on overall computational complexity remains manageable. 

It is worth noting that both contextual filtering and reconstruction enhancement can be treated as plug-and-play tools, meaning they can be selectively disabled in low-resource scenarios, offering flexible trade-offs between performance and speed for practical deployment.

\section{Conclusion}
In this paper, we explore the integration of filtering techniques into the NVC framework. We propose a contextual filtering approach that enhances coding performance by refining contextual information within the coding loop. Additionally, we introduce a reconstruction enhancement module to improve reconstruction quality further. To ensure stable performance, we incorporate an adaptive coding decision mechanism that dynamically determines when to apply these modules, preventing potential degradation while maintaining optimal rate-distortion trade-offs. Experimental results demonstrate that our method achieves competitive performance and often outperforms existing approaches. Future work on techniques such as developing learnable adaptive decision mechanisms and designing lightweight yet highly powerful filtering networks is essential to further enhance filtering performance within NVC frameworks.

\bibliographystyle{ACM-Reference-Format}
\balance
\bibliography{main}

\end{document}